\documentclass[pra,prb,twocolumn,superscriptaddress,floatfix,10pt]{revtex4}
\usepackage{graphicx}
\usepackage{slashbox}
\usepackage{amssymb}
\usepackage{verbatim}

\begin{document}

\title{Electron Density Dependence of in-plane Spin Relaxation Anisotropy \\
in GaAs/AlGaAs Two-Dimensional Electron Gas }

\author{Baoli Liu}
 \affiliation{Beijing National
Laboratory for Condensed Matter Physics,
  Institute of Physics, Chinese Academy of Science,  Beijing, 100080, China}
\author{Hongming Zhao}
\affiliation{Beijing National Laboratory for Condensed Matter
Physics,
  Institute of Physics, Chinese Academy of Science,  Beijing, 100080, China}
\author{Jia Wang}
\affiliation{Beijing National Laboratory for Condensed Matter
Physics,
  Institute of Physics, Chinese Academy of Science,  Beijing, 100080, China}
  \author{Linsheng Lin}
\affiliation{Beijing National Laboratory for Condensed Matter
Physics,
  Institute of Physics, Chinese Academy of Science,  Beijing, 100080, China}
  \author{Wenxin Wang}
\affiliation{Beijing National Laboratory for Condensed Matter
Physics, Institute of Physics, Chinese Academy of Science, Beijing,
100080, China}

\author{Haijun Zhu}
\affiliation{Intelligent Epitaxy Technology, Inc. 1250 E.Collins
Blvd. Richardson, TX 70581, USA}

\author{Dongmin Chen}
 \affiliation{Beijing National Laboratory for Condensed Matter Physics,
  Institute of Physics, Chinese Academy of Science,
  Beijing, 100080, China}

\begin{abstract}
We investigated the spin dynamics of two-dimensional electrons in
(001) GaAs/AlGaAs heterostructure using the time resolved Kerr
rotation technique under a transverse magnetic field. The in-plane
spin lifetime is found to be anisotropic below 150k due to the
interference of Rashba and Dresselhaus spin-orbit coupling
 and D'yakonov-Perel' spin relaxation. The ratio of
in-plane spin lifetimes is measured directly as a function of
temperature and pump power, showing that the electron density in
2DEG channel strongly affects the Rashba spin-orbit coupling.
\end{abstract}
\maketitle

The ability to manipulate the orientation and the relaxation of spin
population in semiconductor two-dimensional (2D) structures via
electrical or optical control is a key step toward building
practical Spintronics devices~\cite{Awschalom2002}. In the prototype
device of spin field effect transistor (FET), proposed by Datta and
Das\cite{Datta 1990}, the Rashba spin-orbit(SO) coupling\cite{Rashba
1984} plays a central role in the controlled rotation of spin via
external electric field in a two-dimensional electron gas (2DEG)
system. In general, the SO coupling includes both Rashba and
Dresselhaus\cite{Dresselhaus 1955} contributions in a realistic
zinc-blende semiconductor 2D structures, and has the undesired
effect of causing spin decoherence in 2DEG at room temperature. The
underlying mechanism is the D'yakonov-Perel' (DP) spin
relaxation\cite{DP,Zakharchenya}, where electron spins randomly
precess about an effective magnetic field resulted from the SO
coupling and thus dependent on the electron's momentum.

Recently, methods for controlling the spin relaxation have been
proposed for a robust spin FET \cite{Loss} and Persistent Spin Helix
(PSH)\cite{Zhang} in 2D structures. The basic idea is to tune the
Rashba and the Dresselhaus terms via proper gating or structure
engineering, so that they have equal strength in (001) 2D structures
or have only the Dresselhaus term in rectangular (110) quantum wells
(QWs). The ability to determine the relative strength of the Rashba
and Dresselhaus terms is, therefore, critical to the design of new
type spin FET and PSH devices. The ratio of Rashba and Dresselhaus
terms at a fixed temperature has been measured previously by several
groups\cite{Golub 2006,Zhang Weber,Ganichev2004}.  The strength of
both Rashba and Dresselhaus SO coupling also have been investigated
by applying a bias at extreme temperature of $\sim$mK
\cite{lu1998,nitta2002}. Theoretically speaking, both Rashba and
Dresselhaus terms produce the in-plane effective magnetic field, and
the interference of these two terms results in the anisotropic
effective magnetic field, and hence a in-plane spin relaxation
anisotropy due to DP mechanism\cite{Wrinkler2003,Golub1999}. Thus
the anisotropy of in-plane spin lifetimes offers a direct
measurement of the relative SO coupling strength. In addition, the
non-equilibrium electrons are generated in a 2DEG system via spin
injection during the operation of realistic spin FET and PSH
devices, and the change of electron density in a 2DEG system can
result in a change of relative SO coupling strength. In this letter,
we use time-resolved Kerr rotation (TRKR) technique to study the
anisotropic in-plane spin lifetime. We address how the electron
density affects the relative strength of the Rashba and Dresselhaus
terms via an elevated temperature and/or pump power. We show that
the electron density in 2DEG channel will strongly affect the Rashba
SO coupling.

The sample studied here consists of GaAs/AlGaAs heterostructure
grown on a (001)-oriented semi-insulating GaAs substrate by
molecular beam epitaxy (MBE). A 500nm GaAs buffer layer first was
grown on the substrate followed by 14nm Al$_{0.24}$Ga$_{0.76}$As
spacer layer, 25nm Al$_{0.24}$Ga$_{0.76}$As Si-doped
4$\times$10$^{18}$cm$^{3}$, and finally a~1~nm GaAs Si-doping cap
layer.~The standard Hall measurement gives the electron
concentration n=6.0$\times$10$^{11}$cm$^{-2}$ at room temperature,
and n=4.5$\times$10$^{11}$cm$^{-2}$ at 150K. We also prepared a GaAs
bulk sample from the same substrate wafer. Two cleaved edges
oriented along [110] and [1$\bar{1}$0] axes were prepared for all
samples. The TRKR experiment was carried out in an Oxford
magneto-optical cryostat supplied with a 7-T split-coil
super-conducting magnet. The sample was excited near normal
incidence with degenerate pump and delayed probe pulses from a
Coherent mode-locked Ti-sapphire laser($\sim$120fs, 76MHz). The
center of the photon energy was tuned for the maximum Kerr rotation
signal for each sample and temperature setting. The laser beams were
focused to a spot size of $\sim$100$\mu$m, and the pump and probe
beams have an average power of 5.0mW and 0.5mW, respectively. The
helicity of linearly polarized pump beam was modulated at 50kHz by a
photoelastic modulator (PEM) for lock-in detection. The circularly
polarized pump pulse incident normal to the sample creates
spin-polarized electrons with the spin vector along the growth
direction of samples. The temporal evolution of the electron spin
was recorded by measuring the Kerr rotation angle
$\theta_{K}$($\Delta$t) of the linearly polarized probe pulse while
sweeping $\Delta$t, which correspond to the net spin component
normal to the sample plane.

\begin{figure}[h]
  \includegraphics[width=7.5cm]{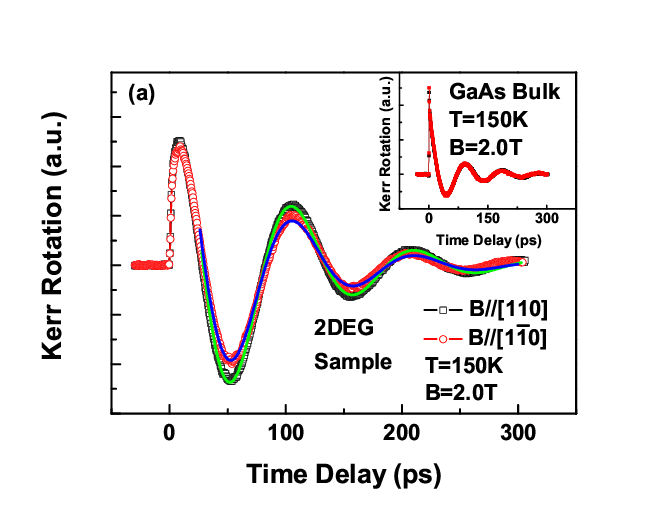}\\
  \includegraphics[width=7.2cm]{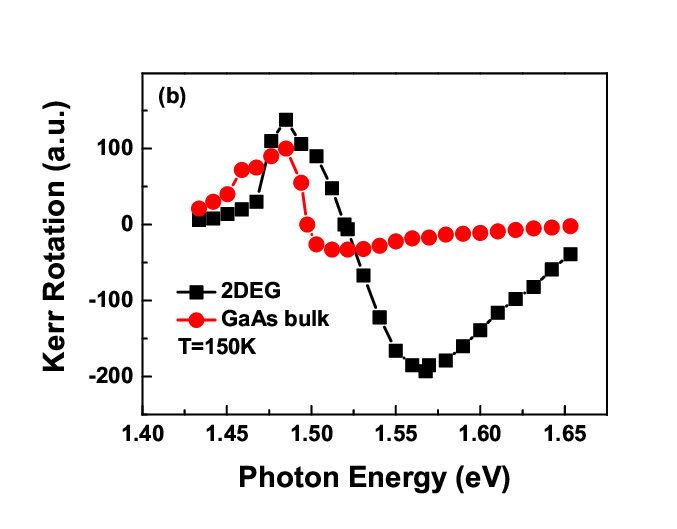}\\
  \caption{\label{fig:TRKR signal}(Color online) (a)TRKR angle $\theta_{K}(\Delta t)$ for 2DEG
  sample measured at B=2.0T and 150K, (b)The spectrum-dependent TRKR angle at $\Delta$t=30ps for
   both 2DEG and GaAs bulk samples. The inset shows the TRKR signal for GaAs bulk sample at B=2.0T and T=150K,}
\end{figure}

Figure\ref{fig:TRKR signal}(a) shows $\theta_{K}$($\Delta$t)
measured at 150K for a 2DEG sample with a in-plane magnetic field of
B=2.0T applied along axes [110] and [1$\bar{1}$0], respectively. The
data show strong oscillations corresponding to the spin precession
with an exponential decay envelope. We found that the amplitude of
oscillation signal of the 2DEG sample with the magnetic field
direction along [110] was larger than that with magnetic field
direction along [1$\bar{1}$0]. The measured signal of the 2DEG
sample may include the contribution of GaAs substrate because the
photon energy of laser is above the band gap of GaAs bulk. To verify
the possible contribution from the GaAs substrate, we measure the
spectrum-dependent Kerr rotation signals at 30 pico-second (ps) time
delay for both 2DEGs and GaAs bulk samples at the same excitation
power with zero magnetic field. As shown in Fig.\ref{fig:TRKR
signal}(b), the signal of the 2DEG sample reaches the maximum at
1.57eV photon energy, which is far from the bandgap of GaAs at 150K.
It is more than 10 times greater than that of pure GaAs bulk sample
at this excitation energy. At the photon energy of 1.48eV, the
signal of both 2DEG and GaAs bulk is almost identical. This energy
corresponds the bandgap of GaAs bulk at 150K. Thus by setting the
excitation beam 1.57eV for our TRKR measurements of the 2DEG sample
as shown in Fig.\ref{fig:TRKR signal}(a), the effect of GaAs
substrate can be safely neglected. Furthermore, we check the TRKR
signal at the different sample locations with magnetic field along a
fixed direction, for instance, [110] axis. The oscillation signals
at all detection positions are essentially the same (data not shown
here), hence we can exclude possible effect due to sample
inhomogeneity in the data reported here. Thus, the data of
Fig.\ref{fig:TRKR signal}(a) indicates that the in-plane spin
lifetimes of the 2DEG sample are different between spins oriented
along [110] and [1$\bar{1}$0].

The spin component normal (S$_{\perp}$) to the 2DEG plane can be
expressed by\cite{oestreich2004,Ohno2005}
\begin{equation}
    S_{\perp}(\Delta t)=S_{0}e^{-(1/\tau_{\perp}+1/\tau_{\parallel})
    \Delta t/2}\cos(\texttt{g}\mu_{B}B\Delta t/\hbar)\label{eq:simple solution}
\end{equation}
where S$_{0}$ is a constant,$\tau_{\parallel}$($\tau_{\bot}$) is the
in-plane (out-of-plane) spin lifetime, g is the electron g factor,
$\mu_{B}$ is the Bohr magneton, $\hbar$ is reduced Plank constant.
The out-of-plane spin lifetime $\tau_{\perp}$ can be obtained by
fitting the experimental data at B=0T with a single exponential
decay, which is around $\sim$110ps at T=150K. Using this value, and
$|$g$|$ and $\tau_{\|}$ as fitting parameters, we obtain good fits
of Eq.~(\ref{eq:simple solution}) to the data in Fig.~\ref{fig:TRKR
signal}(a) shown in green and blue lines, respectively. While
$|$g$|$=0.36 for the both spectra taking with magnetic fields along
[110] and [1$\bar{1}$0] axes, the in-plane spin lifetimes are about
30\% different, however, for spins oriented along [110] and
[1$\bar{1}$0]: $\tau_{\|[110]}$=50ps and $\tau_{\|[1\bar{1}0]}$
=65ps. As a control experiment, we measured the in-plane spin
lifetimes in pure GaAs bulk with magnetic field of B=2.0T along
[110] and [1$\bar{1}$0], respectively, as shown in the inset of
Fig.~\ref{fig:TRKR signal}(a). There is no change of signal
amplitudes, and the same in-plane spin lifetimes of $\tau_{\|}$=80ps
are obtained for both applied magnetic field directions. In bulk
GaAs, the spin splitting originates from the Dresselhaus term, which
results in an isotropic effective magnetic field. As a consequence,
the spin relaxation time in GaAs bulk is isotropic. Thus, we
conclude that the difference in the in-plane spin lifetimes is
related to anisotropic in-plane relaxation in the 2DEG structure.
\begin{figure}[t]
  \includegraphics[width=7.5cm]{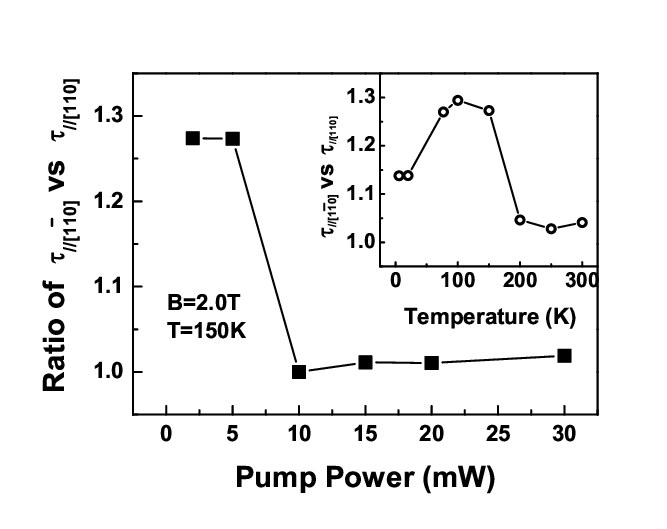}\\
   \caption{\label{fig: ratio of Rashba and Dresselhaus}The ratio of in-plane spin lifetimes measured
  at B=2.0T and T=150K vs pump power. Inset: The ratio of in-plane spin lifetimes
   measured at B=2.0T as a function of temperature.}
\end{figure}

The above observed anisotropy of the in-plane spin relaxation can be
attributed to the interference of Rashba and Dresselhaus SO coupling
in the 2DEG heterostructure\cite{Golub1999,Wrinkler2003}. It should
be enhanced when the strength of both SO coupling is equal, or
reduced when one of SO coupling terms dominates in a 2DEG system.
Thus we can use the change of the anisotropy, i.e. the ratio of
$\tau_{\|[110]}$ and $ \tau_{\|[1\bar{1}0]}$, to monitor the
relative change of SO coupling strength. To test this
interpretation, we tune the relative strength of SO coupling by
manipulating the electron density via pump power and/or temperature
without a gate bias. The advantage of a pure optical control is that
the relationship between the SO coupling and the electron density
can be obtained without change of the band structure due to an
external electrical field. Fig.~\ref{fig: ratio of Rashba and
Dresselhaus} shows the power-dependence of anisotropy with a fixed
probe power (0.5mW) under a magnetic field of B=2.0T and T=150K. It
is evident that the ratio decreases from 1.3 to 1.0 when the pump
power is above 5mW. This means that one of SO coupling overwhelms
the other. At high pump power, only the electron density is changed
in 2DEG which can affect both the Rashba and Dresselhaus SO
coupling\cite{lu1998, nitta2002}. The change of anisotropy in our
experiment, however, indicates a different degree of their response
to the increasing electron density. Since the Rashba effect is the
main source of spin splitting in (001) 2DEG
heterostructure\cite{lu1998}, the trend suggests that the higher
electron density mainly enhance the strength of Rashba term. We
believe that the higher electron density will increase the electric
field of confinement potential and hence the strength of Rashba SO
coupling is increased as well.

Alternatively, the electron density can be raised at elevated sample
temperature in a 2DEG system, and the DP spin relaxation mechanism
dominates in high temperature regime.~The inset of Fig.~\ref{fig:
ratio of Rashba and Dresselhaus}(b)~presents the
temperature-dependent anisotropy at a fixed pump power of $\sim$5mW.
Initially, the ratio increases from 1.15 to 1.3 with temperature up
to 150K. When the temperature is above 200K, this ratio drops to
1.0. The electron density at room temperature is larger than that at
low temperature in our 2DEG system. This observation further
supports that the higher electron density mainly increase the
strength of Rashba term.~Below 77K, the ratio is smaller than that
in high temperature regime (77K$\sim$150K). This means that other
spin relaxation mechanisms such as Elliot and Yafet (EY)\cite{EY}
and Bir, Aronov, and Pikus (BAK)\cite{BAK} will compete with DP
mechanism. Between 77K and 150K, the DP mechanism is dominant. It is
consistent with the results obtained in (110)
QWs\cite{oestreich2004}.

In conclusion, we observed an anisotropy of the in-plane spin
lifetimes in the (001)-orientated GaAs/AlGaAs 2DEG heterostructure,
which exhibits a strong electron density dependence. In particular,
the Rashba term dominates the Dresselhaus term at higher electron
density in the 2DEG sample. These findings could be exploited in
future design of spin-related electronic devices

This work was supported by the Knowledge Innovation Project of the
Chinese Academy of Sciences, the NSFC under the grant No.~10504030,
the Chinese-French PRA project No.~PRA MX06-07.

\end{document}